\documentclass[aps,pre,twocolumn,superscriptaddress,longbibliography,hidelinks]{revtex4-2}

\usepackage{color}
\usepackage{graphicx}
\usepackage{siunitx}
\usepackage{amsmath}
\usepackage{hyperref}

\usepackage{bm}
\usepackage{upgreek}

\begin{document}

\title{Linear theory of viscoelasticity in a generalized hydrodynamic framework}

\author{Andreas M. Menzel}
\email{a.menzel@ovgu.de}
\affiliation{Institut f\"ur Physik, Otto-von-Guericke-Universität Magdeburg, Universitätsplatz 2, 39106 Magdeburg, Germany}

\date{\today}

\begin{abstract}
A generalized hydrodynamic theory that systematically incorporates elasticity and viscoelasticity had been derived about a quarter of a century ago. It is based on a strictly Euler point of view, as is natural for hydrodynamics. We used and adapted this theory particularly in a linear framework. There, it is straightforward to focus on what are the flow and displacement fields in linearized hydrodynamics and elasticity, which provides some advantages. Since this theoretical approach appears not to be as commonly widespread as it deserves to be, we here overview and review the formalism. Specific further focus is on pointing out relations to the commonly known Kelvin-Voigt model and Maxwell model. They are naturally contained within this description. The two limits of perfect long-term elasticity on one hand and long-term flow on the other hand can be represented by adjusting only one parameter. 
\end{abstract}

\maketitle

\section{Introduction}

The generalized theory of hydrodynamics \cite{martin1972unified, pleiner1996hydrodynamics} allows to include additional aspects of material behavior beyond those of ``simple'' liquids \cite{hansen2013theory} into the hydrodynamic framework in a systematic way. Particularly, this is true for the dynamics of quantities that follow conservation laws, such as charges \cite{pleiner1996hydrodynamics}. The dynamics of slowly relaxing macroscopic quantities can be included systematically as well, although the choice of these quantities may be based on intuition and then is phenomenological to a certain degree. In contrast to these,  spontaneously broken symmetries again result in a completely systematic consideration of associated dynamic quantities. An intuitive example is spontaneously broken rotational symmetry in nematic liquid crystals \cite{pleiner1996hydrodynamics, degennes2003physics}. %The same is true for spontaneously broken translational symmetry associated with elasticity. 

In Ref.~\onlinecite{temmen2000convective}, the authors outlined and derived the corresponding generalized hydrodynamic framework that includes elasticity. It also already points out a way of systematic extension to viscoelasticity in general. The term ``viscoelasticity'' is used in different ways, depending on the scientific community. Here, we understand by it any coupling of viscous dynamics and at least partially elastic deformation. This applies to substances that show some long-term flow under mechanical stress applied for a longer time, so that they do not return to their initial shape when the stress is released. Still, they do show some short-term elasticity. Yet, we also call materials viscoelastic that are genuinely elastic in the sense that they do recover their initial shape in this scenario, yet dissipate energy during their motion (``flow'') into the deformed state due to internal friction (``viscosity'').

We note that Ref.~\onlinecite{temmen2000convective} is written in a very compact way. It is therefore quite challenging to access for readers not familiar with this type of generalized hydrodynamic framework, despite the power of the theory. To avoid confusion, we remark that %there was a comment \cite{beris2001comment} associated with Ref.~\onlinecite{temmen2000convective} and a corresponding response \cite{temmen2001temmen}, pointing out that 
the terms ``upper'' and ``lower'' convected derivative had been used in switched convention at some instance in Ref.~\onlinecite{temmen2000convective}, which was remarked later \cite{temmen2001temmen}. 
%The theory was recovered in a different approach a few years later \cite{grmela2002lagrange}. 
There is a slightly more extended description available concerning a few aspects of the theory \cite{pleiner2000structure}, as well as a reconsideration that, amongst others, points out the consistency with the well-known Maxwell model \cite{pleiner2004nonlinear}. 

Still, these works remain complex for readers not familiar with the formalism. Probably, this is why the theory has remained comparatively unnoticed considering the power and potential of the approach. We therefore review it here, providing additional explanations and some further background. Moreover, we focus on the linearized equations for illustration. For many settings, this represents a sufficient description. We further illustrate the connection to the two major models of linear viscoelasticity, namely, the Kelvin-Voigt model and the Maxwell model. Our concept allows for a rephrasing in terms of two vectorial fields instead of the tensorial quantities of strain and stress. Namely, these are the vectorial field of hydrodynamic flow and what is known as the vectorial displacement field in elasticity theory. In this context, the hydrodynamic Euler point of view provides a mind-opening perspective on how to (re-)interpret the elastic displacement field in this framework \cite{puljiz2019memory, richter2021rotating}.

\section{Theoretical framework}
\label{sec:math}

We proceed by reviewing the basic equations of the theory. On our way, we include the input for viscoelastic systems according to the generalized hydrodynamics approach. Where appropriate, we refer to the corresponding parts of the presentation in Ref.~\onlinecite{pleiner2004nonlinear}. 

The first basic equation governing hydrodynamics is the continuity equation stating conservation of mass. We assume incompressibility of our fluid, that is, constant mass density $\rho$. Therefore, in terms of the hydrodynamic flow field $\mathbf{v}(\mathbf{r},t)$, the continuity equation reduces to 
\begin{equation}\label{eq:inkomp}
    \bm{\nabla}\cdot\mathbf{v}=0. 
\end{equation}

Second, conservation of momentum leads to the Navier-Stokes equation in hydrodynamics for a simple liquid. According to Eq.~(2) in Ref.~\onlinecite{pleiner2004nonlinear}, we write it in generalized hydrodynamics as
\begin{equation}\label{eq:vdot}
    \rho\dot{\mathbf{v}} ={} -\bm{\nabla}p+\bm{\nabla}\cdot\bm{\sigma}+\mathbf{f}_\mathrm{b}. 
\end{equation}
In this expression, $p(\mathbf{r},t)$ denotes the pressure field and $\bm{\sigma}(\mathbf{r},t)$ the tensorial stress field. Compared to Ref.~\onlinecite{pleiner2004nonlinear}, the stress tensor $\bm{\sigma}$ is here defined with opposite sign. The sign convention for $\bm{\sigma}$ varies. Here, we choose it to coincide with the sign of the purely elastic stress (see below). We add an external bulk force density field $\mathbf{f}_\mathrm{b}(\mathbf{r},t)$ that may further drive the fluid from outside, for instance, due to gravity. 

To specify the overall stress tensor $\bm{\sigma}$, we have to combine several equations in Ref.~\onlinecite{pleiner2004nonlinear}. First, we turn to the contribution of \textit{viscous stress} that arises from internal friction during motion (``flow''). Since purely translational and purely rotational flows do not show internal friction, it is the symmetrized gradient in velocity that comes into effect here, 
\begin{equation}\label{eq:A}
    \mathbf{A}=\frac{1}{2}\left(\bm{\nabla}\mathbf{v}+(\bm{\nabla}\mathbf{v})^\mathrm{T}\right). 
\end{equation}
In this expression, $^\mathrm{T}$ denotes the transpose. Due to incompressibility, see Eq.~(\ref{eq:inkomp}), its trace vanishes, $A_{kk}=0$, where we use Einstein summation convention throughout. Then, to linear order, the only term that becomes effective in $\bm{\sigma}$ is due to shear viscosity. In the notation of Ref.~\onlinecite{pleiner2004nonlinear}, this contribution reads $\nu_1\mathbf{A}$. $\nu_1$ is the corresponding shear viscosity. [In detail, in Ref.~\onlinecite{pleiner2004nonlinear}, this contribution results from $\sigma_{ij}^{(ph)}$ in Eq.~(3). It is specified in Eq.~(5) as $\sigma_{ij}^{(ph)}=\nu_{ijkl}A_{kl}$. We expand it according to the scheme provided in Eq.~(6) of Ref.~\onlinecite{pleiner2004nonlinear}. Here, we only keep overall linear terms in the stress. That is, all terms of order of the strain tensor $\mathbf{U}$ in Eq.~(6) of Ref.~\onlinecite{pleiner2004nonlinear} are dropped. Then we apply $A_{kk}=0$, so that the term in $\nu_3$ does not contribute.] 

The second contribution to the overall stress tensor $\bm{\sigma}$ results from \textit{elastic stress}. It is formulated in terms of the strain tensor $\mathbf{U}$. Again, we consider only linear contributions for incompressible systems. To linear order, the trace of $\mathbf{U}$ vanishes for incompressible materials, $U_{kk}=0$ \cite{landau2020theory} (see also our remark below). Therefore, the only contribution to $\bm{\sigma}$ resulting from elastic stress to linear order is due to elastic shear deformations. In the notation of Ref.~\onlinecite{pleiner2004nonlinear}, this contribution reads $K_1\mathbf{U}$. $K_1$ is the associated elastic shear modulus. [More precisely, in Ref.~\onlinecite{pleiner2004nonlinear}, the purely elastic stress tensor is termed $\bm{\Psi}$. It is specified according to Eq.~(7) as $\Psi_{ij}=K_{ijkl}U_{kl}$, where $K_{ijkl}$ is expanded as in Eq.~(8) of Ref.~\onlinecite{pleiner2004nonlinear}. From there, to overall linear order in strain and for incompressible systems of $U_{kk}=0$, only the term $K_1\mathbf{U}$ contributes to $\bm{\sigma}$.]

Together, we therefore obtain to linear order and for incompressible systems the expression for the overall stress 
\begin{equation}\label{eq:stress}
    \bm{\sigma} = \nu_1\mathbf{A} + K_1\mathbf{U}.
\end{equation}
The first part results from viscous stress due to motion (``flow''), the second part from elastic deformation. [In Ref.~\onlinecite{pleiner2004nonlinear}, this corresponds to Eq.~(10) after linearization. We recall that we choose the sign of the overall stress tensor $\bm{\sigma}$ oppositely to Ref.~\onlinecite{pleiner2004nonlinear} for consistency with other literature.]

Our third dynamical equation describes the time evolution of strain $\mathbf{U}$. On one hand, deformation is driven by motion (``flow'') into a new state. This is expressed by the symmetrized velocity gradient tensor $\mathbf{A}$ generating strain. In a purely elastic system, this deformation is ``memorized'' in terms of an elastic deformation energy. It increases from zero during deformation of the material out of its undeformed state. Conversely, it reduces back to zero when the elastic material returns into its undeformed state. However, this memory decays over time in a viscoelastic material that shows long-term flow. Due to internal rearrangement, the material tends to adapt to its new shape. Therefore, the memorized strain relaxes over time. The process becomes particularly apparent within the \textit{Euler} perspective, which is the one the hydrodynamic theory is formulated in. To stress this important point once more, we include an extra section below, Sec.~\ref{sec:Euler}. Following the notation of Ref.~\onlinecite{pleiner2004nonlinear}, we denote the additional relaxational contribution that emerges in the dynamic equation for $\mathbf{U}$ as $-\alpha_1K_1\mathbf{U}$. Together, the dynamic equation for the strain tensor $\mathbf{U}$ becomes
\begin{equation}\label{eq:Udot}
    \dot{\mathbf{U}} = \mathbf{A} - \alpha_1K_1\mathbf{U}.
\end{equation}
[More in detail, this equation corresponds to the linearized version of Eq.~(9) in Ref.~\onlinecite{pleiner2004nonlinear}. It includes the relaxational contribution $X_{ij}^{(ph)}=-\alpha_{ijkl}\Psi_{kl}$ from Eq.~(4). Here, $\bm{\Psi}$ is again specified according to Eqs.~(7) and (8) to obtain $K_1\mathbf{U}$ to linear order. Moreover, we expand $\alpha_{ijkl}$ according to Eq.~(6) of Ref.~\onlinecite{pleiner2004nonlinear} to linear order. In both cases, we involve $U_{kk}=0$ for linear incompressibility.]

A very intuitive and classical example for the described process of relaxing strain is provided by creep experiments on polymer melts \cite{strobl1997physics}. Sufficiently long polymer chains show an elevated degree of entanglement. Quickly stretching such a polymeric sample leads to elastic restoring forces that drive it back towards its initial shape directly after stretching. Yet, maintaining the stretched polymeric sample in its stretched state, it adapts to the new shape. This happens due to internal restructuring. Disentanglement of polymer chains over longer time scales occurs \cite{degennes1979scaling, doi1986theory}. The elastic restoring forces decay. Therefore, the system ``loses'' its memory of its initial state. It adopts a new shape and state that it considers as its new undeformed shape and energetic ground state, respectively. Being the new undeformed state, there is no intrinsic strain $\mathbf{U}$ associated with it any more. Thus, the strain $\mathbf{U}$ has relaxed. Since the elastic part of stress and strain are related to each other via elasticity theory, relaxation of elastic stress and strain go hand in hand from this perspective. We employ this relation when we compare to the basic models of linear viscoelasticity.

\section{Comparison with the Kelvin-Voigt model and the Maxwell model}
\label{sec:comparison}

The two basic phenomenological models of linear viscoelasticity are the Kelvin-Voigt model and the Maxwell model. Among them, the Kelvin-Voigt model describes materials of permanent elasticity. They do not show long-term flow. Upon release of imposed external forces, they return to their initial undeformed state. Yet, internal dissipation (``viscosity'') takes place during the dynamical process of motion (``flow'') into and out of a deformed state. Conversely, the Maxwell model allows for terminal flow, that is, permanent remnant changes in shape upon longer-time application of external forces. Restoring elastic stresses that would take the material back to its initial shape decay over time. Both scenarios and phenomenological models are naturally contained in the generalized hydrodynamic theory reviewed in Sec.~\ref{sec:math}, as we demonstrate in the following. 

To identify the Kelvin-Voigt model in Sec.~\ref{sec:math}, we simply solve Eq.~(\ref{eq:Udot}) for the symmetrized gradient velocity tensor $\mathbf{A}$ and insert it into Eq.~(\ref{eq:stress}). In this way, we obtain
\begin{equation}
    \bm{\sigma} = \nu_1\dot{\mathbf{U}} + K_1(1+\alpha_1\nu_1)\mathbf{U}. 
\end{equation}
This equation already has the correct form of the phenomenological Kelvin-Voigt model. Yet, we recall that the Kelvin-Voigt model is typically applied to materials that show long-term elasticity. Therefore, the strain shall not relax as the term with coefficient $\alpha_1$ in Eq.~(\ref{eq:Udot}) would imply. Thus, for such materials, we need to set $\alpha_1=0$. Therefore, the situation corresponding to the Kelvin-Voigt model is well contained as a special case in the generalized hydrodynamic framework revisited in Sec.~\ref{sec:math} \cite{pleiner2004nonlinear} and described by the equation
\begin{equation}\label{eq:Kelvin_Voigt}
    \bm{\sigma} = \nu_1\dot{\mathbf{U}} + K_1\mathbf{U} \qquad (\alpha_1=0).
\end{equation}
Frequently, this type of equation is referred to as constitutive law. We nicely see that both internal dissipation contributes to the stress via the hydrodynamic shear viscosity $\nu_1$ and elasticity contributes via the elastic shear modulus $K_1$. 

Conversely, the link to the Maxwell model has already been pointed out in Ref.~\onlinecite{pleiner2004nonlinear}. We here consider the linearized case, which makes calculations straightforward. For this purpose, we take the time derivative of Eq.~(\ref{eq:stress}), leading to 
\begin{equation}
    \dot{\bm{\sigma}}=\nu_1\dot{\mathbf{A}}+K_1\dot{\mathbf{U}}. 
\end{equation}
For $\dot{\mathbf{U}}$, we insert Eq.~(\ref{eq:Udot}). In the resulting dynamic equation, we replace $\mathbf{U}$ by inferring from Eq.~(\ref{eq:stress}) that $\mathbf{U}=\bm{\sigma}/K_1-\nu_1\mathbf{A}/K_1$. In this way, we reintroduce the stress $\bm{\sigma}$ instead of $\mathbf{U}$. Defining the relaxation time 
\begin{equation}\label{eq:tau1}
    \tau_1=\frac{1}{\alpha_1K_1},
\end{equation} 
the resulting dynamic equation reads
\begin{equation}\label{eq:Maxwell-tau1}
    \tau_1\dot{\bm{\sigma}}+\bm{\sigma} = (\nu_1+\tau_1K_1)\mathbf{A} + \nu_1\tau_1\dot{\mathbf{A}}.
\end{equation}
It is already of the correct form,  but still includes hydrodynamic flows and associated dissipation due to viscosity. To reveal the connection to the Maxwell model, we ignore the role of viscous dissipation. In other words, the associated shear viscosity $\nu_1$ is set to zero. We only focus on internal relaxational restructuring. The latter process is associated with the relaxation parameter $\alpha_1$, see Eq.~(\ref{eq:Udot}). Then, the equation further simplifies to
\begin{equation}\label{eq:maxwell}
    \tau_1\dot{\bm{\sigma}}+\bm{\sigma} = \tau_1K_1\mathbf{A} \qquad (\nu_1=0).
\end{equation}
This dynamic equation was already identified as the Maxwell model in Ref.~\onlinecite{pleiner2004nonlinear}. [It is the linearized form of Eq.~(18) therein.] $\tau_1$, see Eq.~(\ref{eq:Maxwell-tau1}), corresponds to the Maxwell relaxation time. Frequently, in the literature, $\mathbf{A}$ is replaced by the time derivative of a strain tensor, in contrast to our notation and Eq.~(18) in Ref.~\onlinecite{pleiner2004nonlinear}. We return to this point in Sec.~\ref{sec:Euler}. 

Overall, both the Kelvin-Voigt model and the Maxwell model are contained as special cases in the generalized hydrodynamic framework for viscoelastic systems in Sec.~\ref{sec:math}. While the Kelvin-Voigt model describes materials of permanent elasticity ($\alpha_1=0$), the Maxwell model ignores the influence of shear viscosity ($\nu_1=0$) under coupling to motion quantified by the velocity field $\mathbf{v}(\mathbf{r},t)$.

\section{Euler perspective}
\label{sec:Euler}

In the following, we recall the perspective that generalized hydrodynamics takes in describing the physics of complex fluids and materials. It is crucial to stress this point to correctly interpret the corresponding quantities and equations. In contrast to Newtonian mechanics, which we associate with the Lagrange perspective, hydrodynamics takes the Euler perspective. 

The Lagrange perspective is trajectory-based. We follow individual particles or fluid elements over time and describe their trajectories and dynamics in this way. In contrast to this, the Euler perspective is genuinely field-based. Fields, as functions of (fixed) space positions $\mathbf{r}$ and time $t$, evaluate variables and quantities persistently at the fixed positions $\mathbf{r}$ in space. This is true, even if different particles or volume elements pass this position over time. Generalized hydrodynamics adopts the Euler point of view. In this way, we always describe the system in its \textit{present} state in our (fixed) space. 

\begin{figure}
\includegraphics[width=7.5cm]{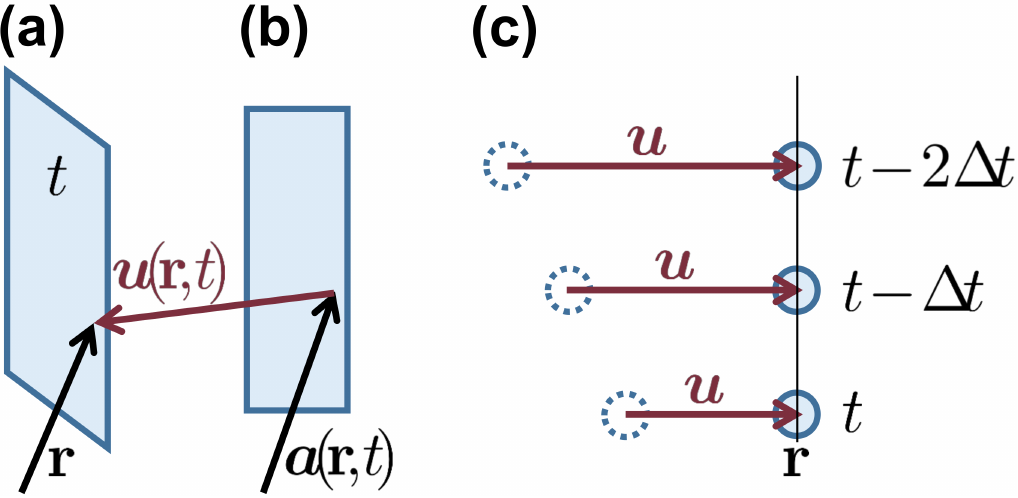}
\caption{In the Euler description associated with generalized hydrodynamics, the material is viewed in its present, possibly deformed state at time $t$ (a). This state is characterized using the fixed space positions $\mathbf{r}$. Instead of recording the trajectories of the material elements starting from an initial position, we note for each position $\mathbf{r}$ in the present, potentially deformed state of the material retrospectively from where the material elements have moved from. For the present state at time $t$ in (a), this defines (b) the field of initial positions $\bm{a}(\mathbf{r},t)$. The field $\bm{u}(\mathbf{r},t)$ links the two. Thus, $\bm{u}(\mathbf{r},t)$ contains the ``memory'' of the initial positions, as viewed from the present state at time $t$ in (a). (c) For viscoelastic materials showing long-time flow, this ``memory'' of initial positions decays over time. If the material were strained, the material element located at $\mathbf{r}$ would tend to relax back to the dotted position at $\mathbf{r}-\bm{u}(\mathbf{r},t)$. Yet, if not perfectly elastic, the material progressively ``forgets'' about the previous state when maintained in the new state. Thus, at time $t$, the memory of where to relax back to has decayed when compared to the memory at earlier times $t-2\Delta t$ and $t-\Delta t$. The (dotted) ``initial'' position to which the material element would move back to approaches its current location $\mathbf{r}$.}
\label{fig:euler}
\end{figure}

For the flow field $\mathbf{v}(\mathbf{r},t)$, this is well known. Conversely, in the context of additional elasticity, the consequences have received less attention so far. 
Always quantifying the system in its present state in a field-based way, we do not consider where the particles or volume elements will move to, but where they have come from. In a sense, this is a backward-oriented perspective. Refs.~\onlinecite{temmen2000convective, pleiner2004nonlinear} termed $\bm{a}(\mathbf{r},t)$ the corresponding field of initial positions of the volume elements, as a function of the present space position $\mathbf{r}$, see Fig.~\ref{fig:euler}(a,b). 

It was argued comprehensibly that the positions $\bm{a}(\mathbf{r},t)$ and $\mathbf{r}$ are associated with different spaces. The first ones form the space of initial positions, while the latter ones constitute the positions in the present space. From this perspective, connecting the elements of these two spaces by defining a displacement field 
\begin{equation}\label{eq:displ}
    \bm{u}(\mathbf{r},t)=\mathbf{r}-\bm{a}(\mathbf{r},t)
\end{equation}
appears mathematically questionable -- how can we define subtractions between elements of different spaces? Indeed, the whole generalized hydrodynamic theory can be formulated purely based on the field $\bm{a}(\mathbf{r},t)$ \cite{temmen2000convective, pleiner2004nonlinear}. Nevertheless, we work with the field $\bm{u}(\mathbf{r},t)$, see Fig.~\ref{fig:euler}(a,b). It is conventional and will not lead to deviations in our linearized framework for isotropic materials. Moreover, it provides an illustrative access to viscoelastic materials, as explained below. 

Indeed, the difference between Lagrange and Euler perspectives already shows up mathematically when we derive the expression for the strain tensor $\mathbf{U}$. It reads $\mathbf{U}=\left[\bm{\nabla}\bm{u}+(\bm{\nabla}\bm{u})^\mathrm{T}\pm(\bm{\nabla}\bm{u})\cdot(\bm{\nabla}\bm{u})^\mathrm{T}\right]/2$. In this expression, ``plus'' results from the Lagrange perspective, viewing the system from its previous state, while ``minus'' is associated with the Euler perspective. The latter describes the system from its present state \cite{chaikin1995principles}. In our linearized presentation, this difference vanishes. Yet the concept of interpreting the associated physical quantities in the appropriate way remains effective. 

Now comes the central conceptual point. If the system is solely described in its present state, it can only know as much as is stored in its present state. In other words, the field $\bm{u}(\mathbf{r},t)$ can only store as much information as is actually available in the system at time $t$. Therefore, if the elastic memory about the initial undeformed state has faded away, for instance, during terminal flow, then this information is lost and all fields characterizing the present state of the system must not contain it. Thus, $\bm{u}(\mathbf{r},t)$, which reflects the elastic memory, must have relaxed to a certain degree, according to the loss of elastic memory and similarly to what is described by the Maxwell model. Still, elastic displacements $\bm{u}(\mathbf{r},t)$ are generated by motion $\mathbf{v}(\mathbf{r},t)$. At the same time, however, if the material is not perfectly elastic, this elastic memory decays over time. The simplest such dynamic relation follows as
\begin{equation}\label{eq:u}
    \dot{\bm{u}} = \mathbf{v}-\gamma\bm{u},
\end{equation}
with one relaxation rate $\gamma$ of the elastic memory. 

Once again, we recall that we are in the Euler frame. %Relaxing $\bm{u}(\mathbf{r},t)$ for $\mathbf{v}(\mathbf{r},t)=\mathbf{0}$ does not mean that the volume elements move back to their initial positions. They do not move (``flow''). Motion of volume elements in the Euler, generalized hydrodynamic picture is always associated with flow $\mathbf{v}(\mathbf{r},t)$. The latter is generated by the stress in Eq.~(\ref{eq:vdot}). Instead, for $\mathbf{v}(\mathbf{r},t)=\mathbf{0}$, the volume elements stay where they are. 
If $\mathbf{v}(\mathbf{r},t)=\mathbf{0}$, then a changing $\mathbf{u}(\mathbf{r},t)$ at position $\mathbf{r}$ according to $\dot{\bm{u}} = -\gamma\bm{u}$ does not imply motion of the volume element at position $\mathbf{r}$. The volume element does not move (``flow''). 
Only the elastically memorized position moves, see Fig.~\ref{fig:euler}(c). It describes the position towards which the volume element would shift back to, if elastic stress were released. This memorized position moves towards the present space position of the volume element, which is $\mathbf{r}$. In this way, relaxation takes place and the system forgets about its initial shape and state. 

The retrospective character of the Euler perspective is reflected in this way. It may not be as widespread as the Lagrange picture. Yet, it has its justification in generalized hydrodynamics and offers a path of (re-)in\-terpreting the role of the field $\bm{u}(\mathbf{r},t)$. 

Conversely, only keeping $\dot{\bm{u}}=\mathbf{v}$ in Eq.~(\ref{eq:u}) and neglecting the contribution $-\gamma\dot{\bm{u}}$, interprets the field $\bm{u}$ as the time-integrated velocity. This picture, being trajectory-based, is associated with the Lagrange perspective. It has its justification when amending the description by constitutive relations. Yet, it takes a different point of view. Its framework is different from the generalized hydrodynamic one, at least if the material allows for terminal flow in the case of $\gamma\neq0$. 

With this background, it becomes clear why we do not simply replace $\mathbf{A}$ by $\dot{\mathbf{U}}$ in Eq.~(\ref{eq:maxwell}), as is frequently found in the literature. This relation results in the Lagrange frame, implying $\mathbf{v}=\dot{\bm{u}}$. $\mathbf{A}$ equal to $\dot{\mathbf{U}}$ would follow directly from inserting $\mathbf{v}=\dot{\bm{u}}$ in the definition of $\mathbf{A}$, see Eq.~(\ref{eq:A}). Then, $\mathbf{A}$ would become identical to the time derivative of the linearized strain tensor
\begin{equation}\label{eq:U}
    \mathbf{U} = \frac{1}{2}\left(\bm{\nabla}\bm{u}+(\bm{\nabla}\bm{u})^\mathrm{T}\right).
\end{equation}
However, in the Euler description, $\mathbf{A}\neq\dot{\mathbf{U}}$ and $\mathbf{v}\neq\dot{\bm{u}}$ in general for $\gamma\neq0$, according to Eq.~(\ref{eq:u}). 

Incompressible, linearly elastic materials are characterized by $U_{kk}=0$ \cite{landau2020theory}. This relation is equivalent to 
\begin{equation}\label{eq:inkompu}
    \bm{\nabla}\cdot\bm{u}=0.    
\end{equation} 
We remark that the time evolution of $\bm{u}$ according to Eq.~(\ref{eq:u}) for linearly (visco)elastic systems maintains $\bm{\nabla}\cdot\bm{u}=0$. This can be inferred by applying the divergence to both sides of Eq.~(\ref{eq:u}). Moreover, we assume $\bm{\nabla}\cdot\bm{u}=0$ at initial time, which is always true for an initially undeformed state. Recalling that $\bm{\nabla}\cdot\mathbf{v}=0$, see Eq.~(\ref{eq:inkomp}), for incompressible systems, Eq.~(\ref{eq:u}) maintains $\bm{\nabla}\cdot\bm{u}=0$.

\section{Linearized description in terms of vectorial quantities}

In a linearized description, it becomes straightforward to reduce the tensorial equations in Sec.~\ref{sec:math} to a vectorial description in terms of the vector fields $\mathbf{v}$ and $\bm{u}$. Our first step is to express $\dot{\mathbf{U}}$ by these vector fields. We take the time derivative of Eq.~(\ref{eq:U}) and then introduce $\dot{\bm{u}}$ as given by Eq.~(\ref{eq:u}). Accordingly, we obtain
\begin{eqnarray}
    \dot{\mathbf{U}} &=& \frac{1}{2}\left(\bm{\nabla}\dot{\bm{u}}+(\bm{\nabla}\dot{\bm{u}})^\mathrm{T}\right) 
    \nonumber\\[.1cm]
    &=&{}
    \frac{1}{2}\left(\bm{\nabla}{\mathbf{v}}+(\bm{\nabla}{\mathbf{v}})^\mathrm{T}\right)
    -\frac{1}{2}\gamma\left(\bm{\nabla}{\bm{u}}+(\bm{\nabla}{\bm{u}})^\mathrm{T}\right)
    \nonumber\\[.1cm]
    &=&{}
    \mathbf{A}-\gamma\mathbf{U}.
    \label{eq:UdotAU}
\end{eqnarray}
We compare this dynamic equation to the time evolution of the strain tensor $\mathbf{U}$ as stated by Eq.~(\ref{eq:Udot}). The form of these two equations is identical. Therefore, the two descriptions are consistent, which confirms our correct interpretation of the field $\bm{u}(\mathbf{r},t)$ in Eq.~(\ref{eq:u}). Moreover, we find an expression for our relaxation parameter $\gamma$. According to Sec.~\ref{sec:Euler}, it quantifies the decay rate of the elastic memory. In terms of the parameters of the generalized hydrodynamic theory, it reads
\begin{equation}\label{eq:gamma}
    \gamma=\alpha_1K_1,
\end{equation}
linking the vectorial and tensorial descriptions. 

Next, we solve Eq.~(\ref{eq:UdotAU}) for $\mathbf{A} = \dot{\mathbf{U}} + \gamma\mathbf{U}$ to calculate from Eq.~(\ref{eq:stress}) the contribution $\bm{\nabla}\cdot\bm{\sigma}$. It emerges in the flow equation, Eq.~(\ref{eq:vdot}). We use Eq.~(\ref{eq:U}) to introduce the field $\bm{u}(\mathbf{r},t)$. Moreover, we again assume incompressibility, which for linear (visco)elastic systems implies both $\bm{\nabla}\cdot\mathbf{v}=0$ and $\bm{\nabla}\cdot\bm{u}=0$, see Eqs.~(\ref{eq:inkomp}) and (\ref{eq:inkompu}), respectively. Accordingly, we obtain
\begin{eqnarray}
    \bm{\nabla}\cdot\bm{\sigma} &=&
    K_1\,\bm{\nabla}\cdot\mathbf{U} + \nu_1\,\bm{\nabla}\cdot\left(\dot{\mathbf{U}}+\gamma\mathbf{U}\right)
    \nonumber\\[.1cm]
    &=&
    (K_1+\nu_1\gamma)\nabla^2\bm{u} + \nu_1\nabla^2\dot{\bm{u}}.
\end{eqnarray}
We introduce this expression into the flow equation, Eq.~(\ref{eq:vdot}). For low Reynolds numbers, the term $\rho\dot{\mathbf{v}}$ on the left-hand side becomes negligible. Then, Eq.~(\ref{eq:vdot}) reduces to
\begin{equation}\label{eq:Stokesu}
    (K_1+\gamma\nu_1)\,\nabla^2\bm{u}+\nu_1\nabla^2\dot{\bm{u}} = \bm{\nabla}p-\mathbf{f}_\mathrm{b}. 
\end{equation}

Ignoring nomenclature, that is, identifying the shear elastic modulus $K_1\equiv\mu$ and the hydrodynamic shear viscosity $\nu_1\equiv\eta$, together with Eqs.~(\ref{eq:u}) and (\ref{eq:gamma}), we therefore recover the dynamic equations evaluated in recent descriptions of viscoelastic materials in terms of the ``memory field'' $\bm{u}(\mathbf{r},t)$ \cite{puljiz2019memory, richter2021rotating}. These characterizations are thus consistent with the linearized generalized hydrodynamic theory for incompressible systems at low Reynolds numbers \cite{temmen2000convective, pleiner2004nonlinear}. 

Finally, we return to the relationship with the phenomenological Kelvin-Voigt model and Maxwell model. As we have demonstrated, the central dynamic equations for the strain tensor $\mathbf{U}$ resulting from the generalized hydrodynamic theory \cite{temmen2000convective, pleiner2004nonlinear} in Eq.~(\ref{eq:Udot}) and from the description based on vector fields \cite{puljiz2019memory, richter2021rotating} in Eq.~(\ref{eq:UdotAU}) are identical. The relation between the two is set by $\gamma=\alpha_1K_1$, as identified in Eq.~(\ref{eq:gamma}). Therefore, also our conclusions drawn in Sec.~\ref{sec:comparison} for the relations to the Kelvin-Voigt model and the Maxwell model carry over to the description based on vector fields, particularly as defined by Eq.~(\ref{eq:u}) for the ``elastic memory field'' $\bm{u}(\mathbf{r},t)$.

The Kelvin-Voigt model results for $\alpha_1=0$, see Eq.~(\ref{eq:Kelvin_Voigt}). This implies that, in the ``memory field''-based description, $\gamma=\alpha_1K_1=0$ in Eqs.~(\ref{eq:u}) and (\ref{eq:Stokesu}). Specifically, in this special case $\dot{\bm{u}}=\mathbf{v}$. Thus, there is no decay of the elastic memory. All initial positions of all volume elements remain memorized over all times. None of their induced displacement by motion $\mathbf{v}$ is forgotten over time. In this way, reversibility of all elastic deformations becomes possible. There is no persistent long-term viscous flow. 

Conversely, the Maxwell model is associated with the case that includes possible relaxation of elastic strain, see the terms $-\alpha_1K_1\mathbf{U}$ in Eq.~(\ref{eq:Udot}) and $-\gamma\mathbf{U}$ in Eq.~(\ref{eq:UdotAU}). It is therefore related to the situation $\gamma>0$. Thus, it implies relaxation of the ``elastic memory field'' according to $\dot{\bm{u}}=\mathbf{v}-\gamma\bm{u}$, see Eq.~(\ref{eq:u}). More precisely, the comparison with the generalized hydrodynamic theory now also provides us with further background of the decay parameter $\gamma$. Combining Eqs.~(\ref{eq:tau1}) and (\ref{eq:gamma}) identifies it as the inverse of the Maxwell relaxation time, 
\begin{equation}
    \gamma=\frac{1}{\tau_1}. 
\end{equation}

\section{Conclusions}
\label{sec:concl}

Generalized hydrodynamics \cite{martin1972unified, pleiner1996hydrodynamics} is a powerful approach to systematically characterize the long-wavelength, low-frequency continuum dynamics of complex materials. It is systematic and therefore provides a benchmark for related descriptions. 

About a quarter of a century ago, it was extended to include elasticity and viscoelasticity of complex fluids and viscoelastic solids \cite{temmen2000convective, pleiner2004nonlinear}. We overview the theory by focusing on a linearized version for illustration and reference. On our way, we provide relations to the basic models of linear viscoelasticity. These are the Kelvin-Voigt model on one hand and the Maxwell model on the other hand. Both are contained in the theory. 

It is the Euler point of view that is central for the generalized hydrodynamic perspective. We address this point in detail. From there, in the linear approach, it is possible to reduce the tensorial description to a vectorial notation. In this context, the displacement field of linear elasticity is (re-)interpreted as an ``elastic memory field'' \cite{puljiz2019memory, richter2021rotating}. Again, this description contains the Kelvin-Voigt model and the Maxwell model. The decay parameter of the elastic memory vanishes in the Kelvin-Voigt case of permanent elasticity. It is identified as the inverse of the relaxation time of the corresponding Maxwell model that facilitates long-term flow. Our presentation is valid for incompressible materials and especially for low-Reynolds-number motion. 

Overall, we may conclude that the linearized description in terms of vectorial notation provides a significantly simplified path to characterize the dynamics of viscoelastic materials. It certainly offers a beneficial framework for situations in which the necessary conditions are met, specifically incompressibility, linear response, and in parts low Reynolds numbers. Potential example situations for its use comprise the behavior of discrete particles in continuous environments \cite{phan1994load, dhont1996introduction, puljiz2017forces, puljiz2019memory, richter2021rotating} and the properties of active viscoelastic suspensions or solids \cite{reinken2025rheologically, reinken2025unified}.

\begin{acknowledgments}
The author thanks the Deutsche Forschungsgemeinschaft (German Research Foundation, DFG) for support through the Heisenberg Grant ME 3571/4-1 as well as grants ME 3571/10-1, ME 3571/11-1, and ME 3571/12-1. 
\end{acknowledgments}

%\bibliography{references.bib}


\end{document}